# Enhancing Software Development Process Using Automated Adaptation of Object Ensembles

Md. Emran, Humaun Kabir, Ziaur Rahman, Nazrul Islam

Department of Information and Communication Technology

Mawlana Bhashani Science and Technology University

Santosh, Tangail-1902, Bangladesh

Email: emranict08@gmail.com, humaunkabirmstu10@gmail.com, zia@iut-dhaka.edu, nazrul.islam@mbstu.ac.bd

*Abstract*— Software development has been changing rapidly. This development process can be influenced through changing developer friendly approaches. We can save time consumption and accelerate the development process if we can automatically guide programmer during software development. There are some approaches that recommended relevant code snippets and API-items to the developer. Some approaches apply general code, searching techniques and some approaches use an online based repository mining strategies. But it gets quite difficult to help programmers when they need particular type conversion problems. More specifically when they want to adapt existing interfaces according to their expectation. One of the familiar triumph to guide developers in such situation is adapting collections and arrays through automated adaptation of object ensembles. But how does it help to a novice developer in real time software development that is not explicitly specified? In this paper, we have developed a system that works as a plugin-tool integrated with a particular Data Mining Integrated environment (DMIE) to recommend relevant interface while they seek for a type conversion situation. We have a mined repository of respective adapter classes and related APIs from where developer, search their query and get their result using the relevant transformer classes. The system that recommends developers titled automated objective ensembles (AOE plugin). From the investigation as we have ever made, we can see that our approach much better than some of the existing approaches.

## I. INTRODUCTION

The Software development process can be enriched by using many processes. Many guidelines are given which influence software development process, especially in the coding stage. Reusing previously completed software repository to enhance the development process is a common phenomenon in the field of Mining Software Repository (MSR). If developers get suggestions in e.g. API recommendations, object usage pattern, class structure or code snippets from the existing projects they might be benefited a lot what they eventually expect while coding. Some of the approaches have integrated web based code, searching in their customized tool like MAC [1] and MAPO [2] before mining the code source abstractions. Although MAC and MAPO are server dependent or online based, by which they are not flexible for a developer. Automated Adaptation of Object Ensembles (AOE) shows a process of adapting the collection frameworks and Arrays, but it is not clear view how to use it as a plug-in tool.

In software development there are different ways available to guide software developers during the development period. A programmer can use programming code in a software system easily by using an automated adaptation of object ensembles. By this process user can find out required data easily. We have used an AOE Plug-in by which a software developer can complete a code in a short time. It takes less time compared to existing approaches. The existing approaches like simply using the Integrated Development Environment (IDE) like Net Beans [3], IntelliJIdea [4], and Eclipse [5], is vulnerable to flaws and it is unable to provide us the required interfaces. In essence, it consumes developers' valuable times. We have investigated between the conventional ways and our AOE plug-in approach to evaluate the efficiency of our proposed tool. The study shows that it is able to accelerate the developer's performance and facilitate less time consuming with decreasing code flaws and errors.

Although in MAC [1], MAPO [2], AOE [6]and many others code repository had several limitations such as server dependent searching, either database based or Internet based where data is not preprocessed according to rules of data mining so these are not friendly for a developer. In our methodology we try to show our repository plugin as an offline repository that is filed based instead of a database and it has special source abstraction technique. Adapter classes are the key point in our approach. In AOE the result depends on the resource of the repository of adapter classes. Recently, a number of works are available to guide developers in the field of software engineering. All we have seen is that our approach is comparatively easier to handle than other existing approaches.

The paper is so far structured in the followings: Section II background and related works of this study. Section III shows that full design and proposed approach. The detailed results and evaluation of this paper is presented in Section IV. Section V concludes with the set of observation and future work of this research.

## II. BACKGROUNDS AND RELATED WORKS

As the software development process can be influenced by using many techniques, so many researchers try to provide a flexible way that can influence the software development process. Need not to say that previously completed software repository technique enhance the development process is a common framework in the field of mining software repository. The developers can be benefited by following the provided suggestions from various recommendations like API recommendations, object usage pattern, class structure or code snippets from the existing projects. Some of the approaches have integrated web based code, searching in their customized tool like MAC [1] and MAPO [2] before mining the code source abstractions.

There are some processes by which a software system is established by code reusing. But in our approach we use a repository of adapter classes and a tool which adapt this code. By using this AOE plug-in which integrates with the IDE, the user can find the required data type easily. There are some existing efforts such as adapting collection and array by using Automated Adaption of Object (AOE)[6]. Some approaches like Code reusing in MAPO [2], better user recommendation using enhancing software development repository, Scenario Based API Recommendation System [7] and others are also used to accelerate the software development process. As we have proposed that if there have been adapter tools [8] which adapt the given interfaces it will be more helpful for the programmers to find the required interfaces. Mining API Usages from Open Source Repositories (MAPO) [2], [3] was one of the first and MAC [1] was one of the updated triumphs to mine API usage pattern. Other recent works called Enhancing Software Development Process (ESDP) [9] where the developers are highly guided by recommendations from a mined repository is also one of our referral works.

One of the best concepts of test-driven reuse showed by Reiss [10] common test-cases issued as input for a component search engine in [11]. Nevertheless, there exist some problems such as license problem and dependency problem. When the user changes the parameter types, then it might need an even more propagated deep adjustment of type changes. The formal and rule-based language is proposed by Kell [12] that was named Cake for automated wrapper generations. The designing used to define interface relations; transformation object structures are possible by applying these rules. We introduced that transformation should happen automatically but it is most overhead for a developer writing mapping rules, basically for an unknown complex object. It overcame Nita and Notkin [13] by providing an approach which concerned with adapting programs to alternative APIs. When the variations among the APIs are small its schema considers not-straightforward structural respect as out of scope, which is the main challenge, is providing transformation. The work showed by Hummel [14] is depending on the Identity Map Pattern from Fowler [15] and identifies the answer about the problems of the Gang of Four adapter pattern. The approach is integrated into another work by Hummel and Atkinson [14] that supplies relaxed-signature matching for primitive data types.

Recently there have done many works to enhance the software development process. There have applied many tools such as adapter generation tool [6]. In this way a user can find the required method easily. As a result a user can save time and solve any problem easily. The software development process is an easy task for a developer.

In the approach we try to overcome the limitations of existing repository tools. We try to provide as an offline repository tool that are file based instead of a database that has special source abstraction. Adapter and transformer classes are the key point in our approach.

## III. PROPOSED APPROACH

In our approach we have developed a plug-in tool that is able to guide software development through suggesting interfaces by using the respective adapter classes. The tool is completely written in java and is executable as a standalone application. It can work with IDE like Net beans [3] and Eclipse [5] as they have the software extensibility.

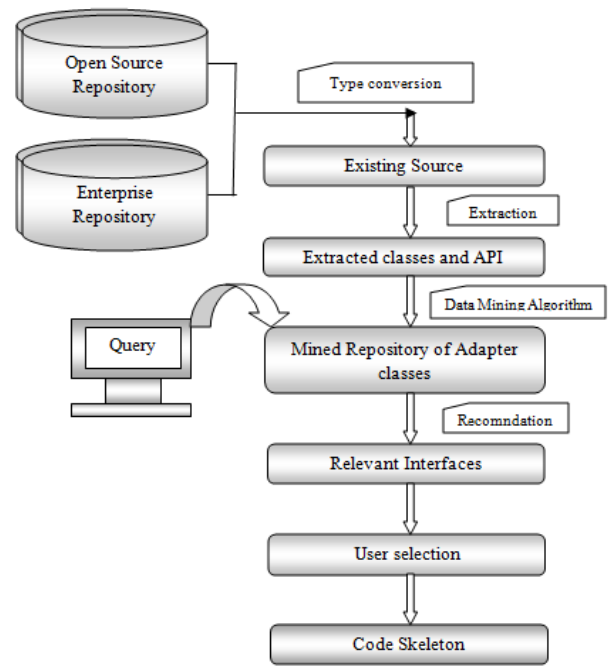

Fig. 1. Automated Object Ensembles (AOE) plugin Framework

Open Source Repository (OSR) is an online software code repository. In this repository many projects problem solving codes are stored. When a programmer stays in online and gets any programming problem, then the programmer can search in this repository for required code. OSP is an Internet based repository. In many software companies, there are stored many projects. It is called the Enterprise Repository.

In repository, there are stored a large number of programs. Many unusual code lines are staying in programs. We take the only important line for a program. By using type conversion we build up Existing Source where projects are stored without unusual code lines. Programming code is stored in Existing Sources. We extract classes of API from these projects which exist in Existing Sources.

Data Mining Algorithm is used to build up a mined repository of Adapter classes. When a programmer searches any classes, the required classes are shown at first. Then we will find related classes. Programmers, search their needed query in the mined repository of Adapter Classes. Then we get relevant interfaces with the help of a transformer and recommendation interfaces. The user selects the required interfaces and gets code skeleton.

Now a day, there has various tools to increase the optimizing capability in software development field; adapter generation tool is one of them. To overcome mismatches on the signature level a good way is to add an adapter that controls message forwarding from one interface to the other. The adapters allow classes to work together that could not otherwise because of incompatible interfaces. It can also be familiar as a "Wrapper" which wraps the incompatible class into the adapter class, where adapter adapt any types of object, method and interface that helps in the programming fields. There exist projects available the newly integrated ability to transform arrays and collections, which can be executed for verifying the adaptation capabilities. This idea is to describe by Hummel [11], where able to generate a random Array List and a sorted Vector instance the helper class Generator is used. When a client uses an automated adapter class that depends on the interfaces that are provided by collection frameworks then the client needs help to use any plugin.

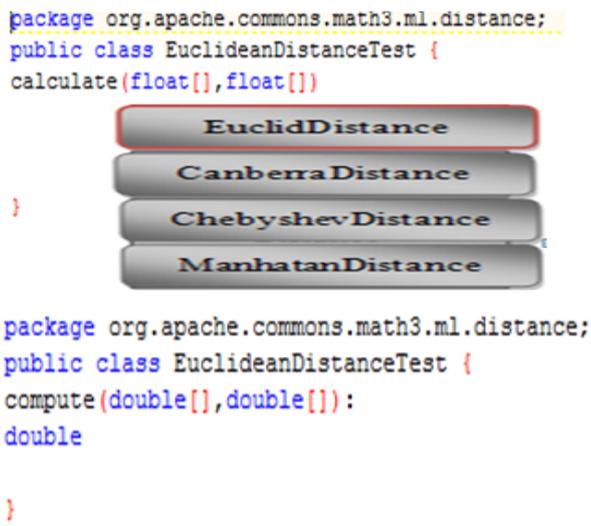

Fig. 2.    Adaptation testing overview

For example, when a programmer writes their code in the IDE like as Net Beans they must write the full code to reach the required result, But when they use AOE plug-in by pressing the right button, then the programmer will be guided by the several adapter classes that are shown in the fig. 2., which is remarked by red color box is chosen. Suppose coder write ,calculate (float [], float []) then AOE plaguing suggest Euclid Distance adoptee adapt that generate in D. Seffert and O. Hummel [7] where adapter depends on transformer that transform provided instance to require instance and vice versa. Suppose a client requires a specific data types, then its search in the adapter fields after matching needed data types client can choose any of them. By finishing the process of adaptation successfully the modified test case was executed here using the final adapter instead of the adapter directly. Nevertheless, test case executed full filly the adapter's transformation capability was verified. For example, in the test case the compute method of the selected class from *org.apache.commons.math3.ml.distance* package is tested. That takes as an input two vectors, showed by an array of type double each, and calculates the distance between them. The distance should be zero, if the same vector is provided as the first and second parameter such as in this example. The public double computes (double [], double []) is the interface of the compute method.

```
package org.apache.commons.math3.ml.distance;
public class EuclideanDistanceTest {
final DistanceMeasure distance = new EuclideanDistance();
public void testZero() {
final double [] a = {0, 1, -2, 3.4, 5, -6.7, 89};
Assert.assertEquals(0, distance. compute(a, a), 0d);
```

The array of type double was replaced with a Link List<Float> after the verification of the original test case executed successfully, where the expected name of the method was changed from compute to calculate. The test case is changed to public void double calculate (Linked List<Float>) for the require interfacing.  The adapter generator overcome a parameter type and the method name mismatch, namely from Linked List<Float> on double [] and calculate in a computer.

   Imagine a client requires one method within two parameters, but there exist more than two parameters in a similar method in the tool then adaptation process solves this complexity. In this paper, we want to describe how easily use collection frameworks as a plug-in tool in software development fields.  The collection framework refers the way of implementing interfaces with the help of several classes that are considered as a supported plaguing tool.

We have seen in the approaches [2] [MAPO] that they have used a code search engine to find the desired item following a search query given by a particular client. Like MAPO this paper also enhances the automated adaptation of object ensembles as a repository tool. In this approach the require source may online or offline repository that contain various kinds of tools. Eventually, the approaches are not only suggesting an adapter class, but also provide its related code implementation to reach the desired goal of a software developer. It works automatically when a client type any code during programing according to their require data types or interfaces. In order to consider the existing paper [6] that shows how to generate an adapter with the help of transforms that can able to solve the matching problems of complex data types. In fig.2. There have given a snapshot of the adapter generation tool. On other existing paper there have no idea about the plugin. But in our paper, we want to use plugin for a user.

A plugin is a software component that adds a specific feature to an existing once. When an application supports plug-in, it enables customization. In our paper we want to provide an adapter plug-in by which any client can complete a task more easily than existing processes. When a client wants to find classes or interfaces, there have given some adapter class options. From these options which is chosen then it finds out the required data by using adapter plug-in.

```
package result;
import java.util.WeakHashMap;

public calss Matrix{

private adaptees Rechteck adaptee;
private static WeakHashMap<adaptees.Rechteck,Matrix>Map=

public Matrix (int param0,int param1,String param2){
    adaptee=new adaptees.Rechteck (param0,param1,param2);
    map.put(adaptee,this);
    }
private Matrix(adaptees.Rechteck adaptee){
    this.adaptee=adaptee;
    map.put(adapte,this);
        }
public adaptees Reckteck getAdaptee(){
    return adaptee;
    }
}
```

Fig. 3. A snapshot 1 of Adapter generation tool (AOE)

## IV. EVALUATION

There have given a guideline for a programmer to enhance the software development process. An automated adaptation of object ensembles is a process by which any software developer can find the required data easily. There have given some comparison by which we will understand which process is better than another.

### a. Environmental setup

In this development process, there have used a repository of adapter classes. How much healthier this repository the development process is more easy. If our adapter class repository is enriched, a user can find the required interfaces in a short time and easily. There have given same adapter class name such as:

| SL | Adapter Classes | Line of Code (LOC) |
|----|-----------------|--------------------|
| A | EuclidDistance | 17 |
| B | CanberraDistance | 11 |
| C | ChebyshevDistance | 15 |
| D | ManhattanDistance | 14 |
| E | OnewayAnova | 16 |

There used an adapter generation tool. Some object oriented language is used. There needs an adapter by which adaptation code is plagued with user. There have given an adapter generation tool such as adaptation tool.zip [6].

### b. Time complexity

Time is an important thing when a program is solved. By which a program is solved quickly this process is better than other. When a programmer solves a program in IDE (Net Beans, eclipse) there have needed some times such as 40sec.

TABLE I. TIME COMPLEXITY

| Task | IDE with AOE plug-in(no of solved problem) | NetBeans(no of solved problem) |
|------|--------------------------------------------|-------------------------------|
| T1 | 60 sec | 120 sec |
| T2 | 130 sec | 200 sec |
| T3 | 150 sec | 310 sec |
| T4 | 120 sec | 360 sec |
| T5 | 140 sec | 400 sec |

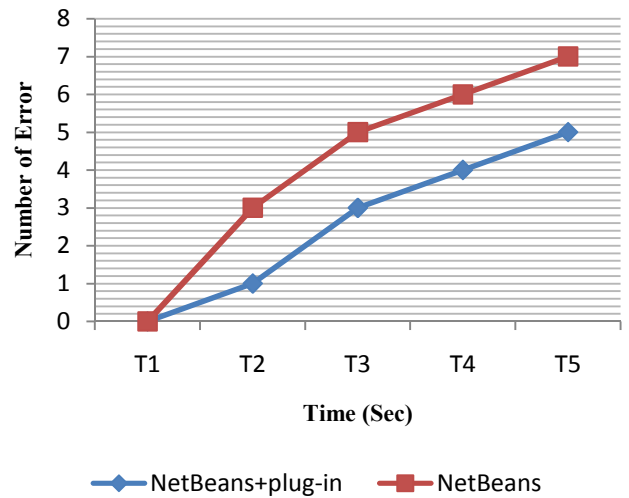

Fig. 4. Time complexity between NetBeans and NetBeans+plug-in But the same programmer solves this same problem in net beans, but there have plugged-in AOE. As a result, we see that at this time the programmer can solve this problem in 0.20sec. When AOE is plugged-in, they have saved 20sec. There have given a table of how much program is solved without AOE plugged-in and within AOE plugged-in.

c. *Error Vulnerability*

In TABLE II we see that after the total time in IDE with plague-in there have found 13 errors, but at the same time problem solving only in IDE there have found out 21 errors.

TABLE II. Error possibility

| Time | NetBeans+plug-in | NeatBeans |
|---|---|---|
| T1 | 0 | 0 |
| T2 | 1 | 3 |
| T3 | 3 | 5 |
| T4 | 4 | 6 |
| T5 | 5 | 7 |
| Total error | =13 | =21 |

As a result, we understand that when the plugin is used error rate is low. So this process is better than another.

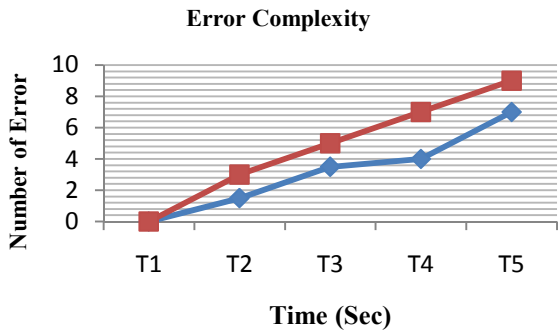

Fig. 5. Error comparison between net beans and NetBeans+plug-in

d. *Thread of the evaluation*

Everything has a limitation. There has some limitation of evaluation. This guideline for software development process is more effective. The evaluation is observed at the same time and same experiment. Such as

a. By having more AOE adapter classes in the repository a user finds more accurate data.

b. A program is evaluated by the same user.
c. It is a lengthy process to plug in a user in an adapter class repository.
d. A user cannot find the mining data.

The results observed in the empirical study may not be applicable to the programming tasks other than those considered in the study, being a threat to the external validity. If the tasks mentioned out there in the study change the results may also be changed. Before we start our evaluation the team members are well trained. The receiving capacity of team members may vary. So the learning curve of these numbers may affect the results.

Within many problems this guideline for a programmer is more effective to develop software process. By this process a user can find any data very quickly. It will keep an important role to enhance the software development process.

V. CONCLUSION

The AOE plug-in approach is more enriched than any other Existing approaches. The approach is able to find out any required data easily and there have needed less time than any other approach. A user finds a data in adapter classes than the AOE plug-in give the required data to the client. The process which we have provided in this paper is different from any other existing process. That is able to enhance the SDP recently.

This paper provided a guideline by which a user can get the required data easily and it is comfortable to use. But these data are not mined. In Future there have used data mining algorithm and find out mined data to enhance software development process.

REFERENCES

[1] Hsu, Sheng-Kuei, and Shi-Jen Lin. "MACs: Mining API code snippets for code reuse." *Expert Systems with Applications* 38.6, p. 7291-7301, 2011.
[2] Xie, Tao, and Jian Pei. "MAPO: Mining API usages from open source repositories." *Proceedings of the 2006 international workshop on Mining software repositories*. ACM, p. 54-57, 2006.
[3] NetBeans plug-in(2015),Retrieved from: https:// www.plugins.netbeans./ [Accessed June 28, 2015 ]
[4] Jet Brain, (2015), Retrieved from: https:// www.jetbrains.com/idea/[Accessed" June 25, 2015 ]
[5] Genuitec,(2015),Retrievedfrom:http://www.genuitec.com/myeclips
[6] Seiffert, Dominic, and Oliver Hummel. "Adapting Collections and Arrays: Another Step towards the Automated Adaptation of Object Ensembles." *Software Reuse for Dynamic Systems in the Cloud and Beyond.* Springer, pp. 348-363, 2014
[7] Shahnewaz, "A Scenerio based API recommendation system using syntax and semantics of client source code". Master's Thesis, Department of CSE, Islamic University of Technology, OIC, Gazipur-1704, Bangladesh. April, 2014
[8] http: // oliverhummel. com / adaptation / tool.zip [Accessed July 29, 2015 ]

[9] Rahman, Z., Hasan, K., *"Enhancing Software Development using Data Mining Integrated Environment,"* Master's Thesis of Islamic University of Technology, Dhaka, pp. 1-75, June 2012.

[10] Reiss, S.P.: Semantics-based code search. *In: IEEE 31st International Conference on Software Engineering, ICSE* 2009, pp. 243–253 (2009)

[11] Hummel, O., Janjic, W.: Test-Driven Reuse: Key to Improving Precision of search Engins for Software Reuse. *In: Finding Source Code on the Web for Remix and Reuse*, pp. 227–250. Springer(2013)

[12] Kell, S.: Component adaptation and assembly using interface relations. *In: Proceedings of the ACM International Conference on ObjectOriented Programming Systems Languagesand Application*, OOPSLA 2010, pp. 322-340 , 2010.

[13] Nita, M., Notkin, D.: Using twinning to adapt programs to alternative apis. In: Proceedings of the *32nd ACM/IEEE International* Conference on Software Engineering, ICSE 2010, vol. 1, pp. 205–214 ,2010

[14] Hummel, O., & Atkinson, C. "Automated creation and assessment of component adapters with the test cases*" In Comonent-Based Software Engineering* (pp. 166-181). Springer Berlin Heidelberg.

[15] Janjic, W., Atkinson, C.: Leveraging software search and reuse with automated software adaptation. *In: 2012 ICSE Workshop on Search Driven Development - Users, Infrastructure, Tools and Evaluation*, pp. 23-26 , 2012